\documentclass[rnote]{aa} 
\usepackage{graphicx}
\usepackage{natbib}
\usepackage{txfonts}
\bibpunct{(}{)}{;}{a}{}{,}
\begin{document}

  \title{Comparison of theoretical white dwarf cooling timescales}
  \author{Maurizio Salaris\inst{1},  
          Leandro G. Althaus\inst{2,3} \and 
          Enrique Garc\'ia-Berro \inst{4,5}}
\institute{Astrophysics Research Institute, Liverpool John Moores University, 
           Liverpool Science Park, 146 Brownlow Hill, 
           IC2 Building, Liverpool L3 5RF, UK\\
           \email{M.Salaris@ljmu.ac.uk}     
           \and
           Grupo de Evoluci\'on Estelar y Pulsaciones. 
           Facultad de Ciencias Astron\'omicas y Geof\'{\i}sicas, 
           Universidad Nacional de La Plata, 
           Paseo del Bosque s/n, 
           1900 La Plata, 
           Argentina\
           \and
           Member of CONICET, Argentina\\
           \email{althaus@fcaglp.unlp.edu.ar}     
           \and
           Departament de F\'\i sica Aplicada,
           Universitat Polit\`ecnica de Catalunya,
           c/Esteve Terrades 5, 
           08860 Castelldefels, Spain\\
           \email{enrique.garcia-berro@upc.edu}
           \and       
           Institute for Space  Studies of Catalonia,
           c/Gran Capit\`a 2--4, Edif. Nexus 104,   
           08034  Barcelona,  Spain\\
}
  \abstract
  {An accurate assessment of white dwarf cooling times is paramount to
    place white dwarf cosmochronology  of Galactic populations on more
    solid grounds. This issue is  particularly relevant in view of the
    enhanced   observational  capabilities   provided   by  the   next
    generation  of Extremely  Large Telescopes,  that will  offer more
    avenues to employ white dwarfs as probes of Galactic evolution and
    test-beds of fundamental physics.}
  {We estimate for the first  time the consistency of results obtained 
    from independent  evolutionary codes  for white dwarf  models with
    fixed mass  and chemical stratification, {\sl when  the same input
    physics is employed in the calculations.}}
  {We compute and compare cooling  times obtained from two independent
    and  widely  used stellar  evolution  codes  -- BaSTI  and  LPCODE
    evolutionary codes  -- using exactly  the same input  physics, for
    $0.55\, M_{\sun}$  white dwarf  models with  both pure  carbon and
    uniform  carbon-oxygen  (50/50 mass  fractions)  core  , and  pure
    hydrogen layers  with mass  fraction $q_{\rm  H} =  10^{-4} M_{\rm
    WD}$ on top of a pure helium  buffer of mass $q_{\rm He} = 10^{-2}
    M_{\rm WD}$.}
  {Using  the same  radiative and  conductive  opacities, photospheric
    boundary conditions,  neutrino energy  loss rates and  equation of
    state, cooling times from the two codes agree within $\sim 2\%$ at
    all  luminosities,  except when  $\log(L/L_{\sun})  > -1.5$  where
    differences  up to  $\sim 8\%$  do  appear, due  to the  different
    thermal structure of the first white dwarf converged models at the
    beginning of the cooling sequence. This agreement is true for both
    pure carbon and  uniform carbon-oxygen stratification core models,
    and also when  the release of latent heat  and carbon-oxygen phase
    separation  are considered. We have also determined quantitatively  
    and explained     
    the effect of varying equation of state, low-temperature radiative opacities 
    and electron conduction opacities in our calculations, }
  {We have assessed  for the first time the  maximum possible accuracy
    in the current  estimates of white dwarf  cooling times, resulting
    only from  the different implementations of  the stellar evolution
    equations and homogeneous input physics in two independent stellar
    evolution  codes.    This  accuracy  amounts  to   $\sim  2\%$  at
    luminosities   lower  than   $\log(L/L_{\sun})\sim  -1.5$.    This
    difference is smaller than the  uncertainties in cooling times due
    to  the   present  uncertainties  in  the   white  dwarf  chemical
    stratification.   Finally, we  extend  the scope  of  our work  by
    providing tabulations  of our  cooling sequences and  the required
    input physics,  that can be used  as a comparison test  of cooling
    times obtained from other white dwarf evolutionary codes.}
\authorrunning{M. Salaris et al.}
\maketitle


\section{Introduction}

During the last  two decades white dwarf observations  and theory have
improved to a level that has  finally made possible to start employing
white dwarfs as credible astrophysical  tools -- see \cite{review} for
a recent  review.  A  detailed assessment of  the accuracy  of current
estimates  of  white  dwarf  cooling rates  is  therefore  a  pressing
necessity,  for  carbon-oxygen  white dwarfs  are  increasingly  being
employed  to   constrain  the  age   and  past  history   of  Galactic
populations,  including the  solar  neighbourhood,  open and  globular
clusters    --    see,     i.e.,    \citet{winget},    \citet{Nature},
\citet{hansen2},   \citet{bedin1},  \citet{winget3},   \citet{bedin2},
\citet{garciaberro},  and   references  therein.    Also,  theoretical
estimates of white dwarf cooling  rates are routinely adopted to place
constraints on the properties of  neutrinos, exotic particles and dark
matter  candidates \citep{freese,  primero, winget2,  bertone, isern2,
corsico:12} and alternative theories of  gravity -- see, for instance,
\citet{gdot1},  \citet{gdot3},  and  \citet{gdot2}.   These  types  of
investigations  all  demand an  accurate  calculation  of white  dwarf
cooling  models.  This,  in  turn, requires  a  detailed and  accurate
knowledge of the main physical  processes that affect the evolution of
white  dwarfs, and  the initial  chemical stratification  for a  given
value of the white dwarf mass.

There  have  been a  few  recent  theoretical  studies to  assess  the
sensitivity of  the predicted  cooling times  to uncertainties  in the
model  core   and  envelope   chemical  stratification   and  electron
conduction   opacities,   \citep{hansen,   pradamoroni,   salarisproc,
salarisWD2}  and   photospheric  boundary   conditions  \citep{hansen,
salarisWD, rohrmann}.  However, there is no modern systematic study of
the  effect  of  employing  different equations  of  state  (EOS)  and
radiative opacities,  especially the less  established low-temperature
opacities in  cool white dwarfs.   Besides these potential  sources of
uncertainties,  it  is even  more  pressing  the  need to  assess  the
consistency of  results obtained  from independent  evolutionary codes
for white  dwarf models  with a  fixed mass,  adopting the  same input
physics and chemical stratification.  Differences determined from this
class of  comparisons represent the  maximum possible accuracy  in the
current estimates of white dwarf cooling times, determined only by the
different implementations of the stellar evolution equations and input
physics.  Assessing  the consistency  of results of  independent white
dwarf stellar codes becomes absolutely  necessary to place white dwarf
cosmochronology on  solid grounds.  This  is even more  important when
considering that the next generation  of Extremely Large Telescopes --
i.e., the European-Extremely Large Telescope (E-ELT), the Thirty Meter
Telescope (TMT) and the Giant Magellan  Telescope (GMT) -- and the new
generation of  astrometric satellites  -- like Gaia  -- will  open new
avenues to exploit the potential of white dwarfs as probes of Galactic
evolution  and  fundamental physics  --  see,  i.e., \citet{bono}  and
\citet{gaia}, respectively.  Self-consistent  comparisons of this type
have never been performed.  Similar  tests discussed previously in the
literature \citep{wvh, hp} are not completely consistent, in the sense
that the different codes compared  were not employing exactly the same
input physics, even  though in one case \citep{wvh} the  effect of the
different input  physics adopted in  the models available at  the time
was estimated, to somewhat reduce all calculations to approximately to
the same physics setup.

For  these reasons, we  present here  the first  fully self-consistent
comparison  of cooling  times obtained  using exactly  the  same input
physics  from two  independent  evolutionary codes  whose white  dwarf
calculations  have been widely  employed in  the literature:  The {\tt
LPCODE} evolutionary code  -- see \citet{althaus10} and \citet{renedo}
for recent references -- and  the {\tt BaSTI} evolutionary code -- see
\citet{salarisWD2} and  references therein.  This is done  for a white
dwarf  model with  fixed  chemical stratification.   This approach  is
similar   to   the  crucial   tests   performed   in   the  field   of
asteroseismology --  see, i.e., \citet{marconi},  and \citet{lebreton}
-- where  the  same  physics  input  is adopted  to  compare  internal
structure, evolution  and seismic properties of stellar  models.  As a
byproduct of  these comparisons we also determine  a rigorous estimate
of the effect of  varying the low-temperature radiative opacities, EOS
and   electron  conduction   opacities  amongst   currently  available
tabulations.

  To inspire other members of  the white dwarf community to compare
  results from their evolutionary codes  with this set of calculations
  (thus  broadening the  scope  of this  investigation by  considering
  additional codes) we will make  available tables with the results of
  our reference calculations discussed in  the paper, and the physical
  ingredients  adopted in  these  calculations that  are not  publicly
  available.   This will  be the  first paper  in a  series aiming  to
  assess   comprehensively   the    uncertainties   of   white   dwarf
  cosmochronology. It will be followed by an analysis of the effect of
  standard assumptions in the white  dwarf calculations, like using --
  as   customary  --   pure  H   and/or  He   buffers  with   chemical
  discontinuities   at   the   boundaries   vs.   diffusive   chemical
  transitions,   and  --   relevant  for   bright  white   dwarfs,  as
  demonstrated in this  paper -- comparison of  models evolved through
  the thermally pulsing phase with  models started artificially at the
  top of the  cooling sequence. The final paper will  try to establish
  the  current  most  accurate  physics inputs  (e.g.,  EOS,  boundary
  conditions,   and   opacities)    for   white   dwarf   evolutionary
  calculations.  The outline  of this paper is  as follows. Section~2
describes briefly the codes and their standard assumptions about input
physics,  while  Sect.~3  presents  comparisons of  cooling  times  by
altering step-by-step the model physics  until all inputs are the same
in both sets of calculations.  Conclusions close the paper.

\section{Calculations and comparisons}
\label{cal}

Details  about the  input physics and  numerical solution  of the
  stellar  structure  equations  in  the {\tt  BaSTI}  code  and  {\tt
    LPCODE}, mesh distribution  of the models, opacity  and EOS tables
  interpolations,  are  given  in   the  online  appendix.   We  have
calculated two sets  of cooling models, both for a  white dwarf with a
total mass $M=0.55\, M_{\sun}$, and an envelope consisting of a pure H
layer with mass fraction $q_{\rm H} =  10^{-4} M_{\rm WD}$ on top of a
pure He layer of  mass $q_{\rm He} = 10^{-2} M_{\rm  WD}$ .  The first
group  of calculations  envisages  a  pure carbon  core.   As a  first
baseline set  we have employed  the standard input physics  choices of
the two codes,  the Eddington $T(\tau)$ relation  for the photospheric
boundary condition  and -- to isolate  the effect of just  basic input
physics   --  no   latent  heat   release  upon   crystallization  was
considered. The different choices  of physics inputs in  the two
  codes  are the  low temperature  opacities (below  8000~K), electron
  conduction opacities and EOS (see appendix).

\begin{figure}
\centering
\includegraphics[width=\columnwidth]{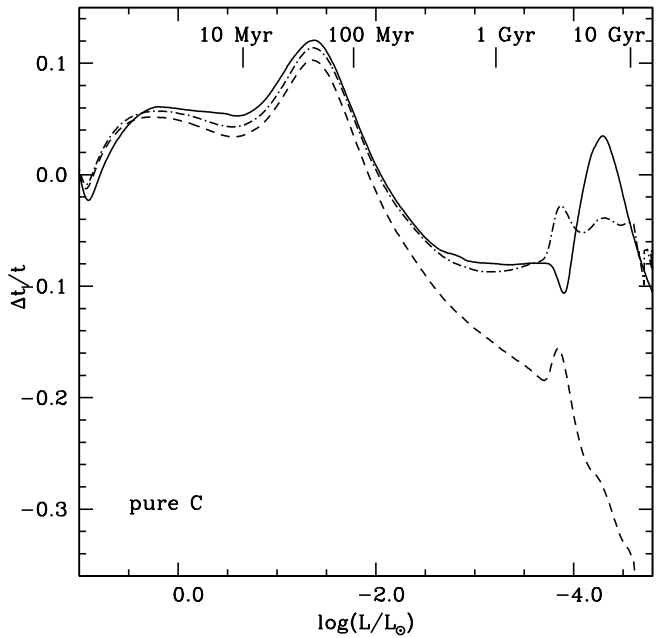}
\caption{Relative  age  difference $\Delta  t/t=(t_{\rm  BaSTI}-t_{\rm  
  LPCODE})/t_{\rm  BaSTI}$ as  a function  of the  luminosity for  the
  $0.55\, M_{\sun}$  carbon-core sequence. The case  with the standard
  input physics of the two codes (detailed in Sect.~2) is displayed as
  a  solid  line.  No  latent  heat  release upon  crystallization  is
  included, and  an Eddington $T(\tau)$  relation is employed  for the
  outer boundary conditions.  Selected ages from the {\tt BaSTI} model
  are  also marked  at  representative  luminosities. Dash-dotted  and
  dashed lines display  additional comparisons by changing  the EOS in
  the model calculations (see text for details).}
\label{FigpureCstd}
\end{figure}

Cooling times for these calculations --  the origin of the cooling age
is set to zero at $\log(L/L_{\sun})=1.1$ in all calculations discussed
in this paper -- are compared in Fig.~\ref{FigpureCstd}.  This diagram
displays  with  a  solid  line the  relative  age  difference  $\Delta
t=(t_{\rm BaSTI}-t_{\rm  LPCODE})/t_{\rm BaSTI}$ as a  function of the
luminosity of  the white dwarf.  Ages  from the {\tt BaSTI}  model are
also marked  at representative luminosities\footnote{The  exact values
  of  these  ages will  change  in  the additional  calculations  that
  follow, but  the order  of magnitude  of the  age at  these selected
  luminosities  will  stay  the  same in  all  calculations  discussed
  here.}.   As can  be  observed, {\tt  BaSTI}  ages appear  typically
larger at luminosities above  $\log(L/L_{\sun})\sim -2.0$, and smaller
at lower  luminosities, apart from the  spike at $\log(L/L_{\sun})\sim
-4.2$.  In  spite of some different  physics inputs in the  two codes,
age differences are within just about $\pm$10\%, at fixed luminosity.

We have  then proceeded  to calculate additional  sets of  pure carbon
models (still  no latent heat release at  crystallization) changing --
one  at a  time --  the inconsistent  physics input  according  to the
following steps:

\begin{enumerate}
\item  Calculations with  both the  {\tt BaSTI}  and the  {\tt LPCODE}
  evolutionary codes  employing the \citet{MM79} EOS  everywhere, that
  is, in both the core and the envelope. We have used here exactly the
  same EOS numerical  routine in both codes.  We use  this EOS because
  the routine is easy to implement  and covers the entire structure of
  the white  dwarf, thus simplifying  the replacement of  the standard
  EOS choices  in both  codes.  Comparisons of  cooling times  at this
  step  cancel  the  effect  of  using a  different  EOS  in  the  two
  calculations.
\item Calculation of {\tt  BaSTI} models with the previously mentioned
  EOS and  electron conduction opacities by \citet{Caea07},  as in the
  {\tt LPCODE}  calculations. The same numerical  routine to calculate
  the electron conduction opacities is  used, but the two codes employ
  different sets  of total (radiative plus  conduction) opacity tables
  and  different interpolation  schemes.  Comparisons  of  {\tt BaSTI}
  model cooling  times at this step  with {\tt LPCODE}  results at the
  previous step cancel the effect  of the difference in the conduction
  opacities between the two calculations.
\item Calculations  of both {\tt  LPCODE} and {\tt BaSTI}  models with
  the   EOS  of   \citet{MM79},  electron   conduction   opacities  by
  \citet{Caea07}, employing now photospheric boundary conditions taken
  at  an  optical  depth  $\tau=25$  from  the  model  atmospheres  by
  \citet{rohrmann}  when $T_{\rm  eff} <10000$~K.   The same  table of
  photospheric  boundary  conditions  is  used in  both  calculations.
  Comparisons  of cooling  times  at this  step  cancel the  remaining
  effect  of using  different  low temperature  opacities  in the  two
  calculations,  for in  cool white  dwarf models  and at  the optical
  depths  where  the boundary  condition  is  fixed,  the envelope  is
  convective and largely adiabatic. This makes a detailed knowledge of
  the low-temperature radiative opacity much less relevant.
\end{enumerate}

\begin{figure}
\centering
\includegraphics[width=\columnwidth]{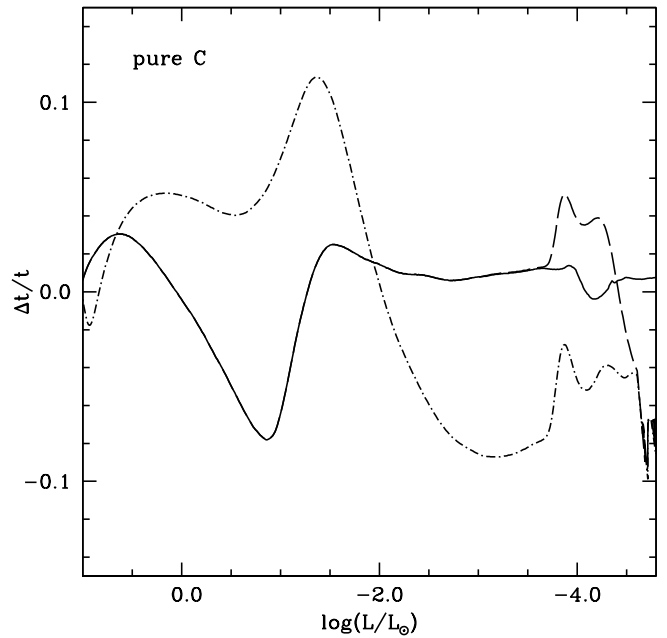}
\caption{Relative  age  difference $\Delta  t/t=(t_{\rm  BaSTI}-t_{\rm
  LPCODE})/t_{\rm BaSTI}$ as a  function of luminosity for carbon-core
  sequences.   No   latent  heat   release  upon   crystallization  is
  assumed. Dash-dotted line, dashed line, and solid line correspond to
  different input  physics as specified  at steps 1,  2, and 3  in the
  text. In  particular, at step 3,  the input physics assumed  in both
  codes is exactly the same (see text for details).}
\label{FigpureC}
\end{figure}

The results of  the comparisons at step 1 (displayed  as a dash-dotted
line),   2  (dashed   line)  and   3   (solid  line)   are  shown   in
Fig.~\ref{FigpureC}.  We start discussing first the effect of changing
the EOS,  by recalling that  in the  standard input physics  case {\tt
BaSTI} and {\tt LPCODE} models shared  the same EOS at high densities,
but  employed a  different EOS  in  the low-density  regime.  To  this
purpose, in addition to the relative age differences obtained with the
standard   input   physics   of    the   two   codes   (solid   line),
Fig.~\ref{FigpureCstd}  displays  also $\Delta  t/t  $  for the  input
physics   at   step   1   (dash-dotted  line,   as   shown   also   in
Fig.~\ref{FigpureC}) and  the age  difference between the  {\tt BaSTI}
calculations at step 1 and the {\tt LPCODE} calculations with standard
physics  (short-dashed   line).   A   comparison  of  the   solid  and
short-dashed line  -- that display $\Delta  t/t $ with respect  to the
same   reference   cooling  track   --   shows   how,  starting   from
$\log(L/L_{\sun})\sim -2.0$, the  \citet{MM79} EOS causes increasingly
and substantially  shorter cooling timescales  compared to the  use of
the EOS by \citet{segretain} at  high densities, and \citet{Saea95} in
the envelope. The main reason for these differences is the lower value
of the specific heat for the  carbon core, that reaches differences of
$\sim$40-50\%  at  the centre  of  the  faintest  models.  If  we  now
consider  the dash-dotted  line, that  shows the  comparison with  the
input  physics  of  step  1,  e.g., with  the  same  \citet{MM79}  EOS
everywhere in both  sets of calculations, $\Delta t/t $  moves back to
be very close to the case with standard physics.

By recalling that for the standard  case (solid line) the EOS employed
in  the two  sets  of models  was different  only  in the  low-density
regime,  and  that  at  step  1 (dash-dotted  line)  it  is  the  same
everywhere,  the  comparison  of  $\Delta t/t$  for  these  two  cases
suggests that  the difference between \citet{Saea95}  and \citet{MM79}
EOS  at  low-densities  --  e.g.  differences  of  the  specific  heat
increasing  with decreasing  luminosities  up to  $\sim \pm$50\%,  and
smaller  differences  of the  specific  heat  of  the order  of  $\sim
\pm$10-20\% -- has a very small  effect on the cooling times, at least
down to $\log(L/L_{\sun})\sim  -4.0$. At these low  luminosities it is
the onset of convective coupling \citep{dm:90, fbb:01} that causes the
different behaviour of $\Delta t/t$  between the two cases. Convective
coupling occurs when the base  of the convective envelope reaches into
degenerate layers (within the  hydrogen envelope) coupling the surface
with  the  degenerate interior,  and  increasing  the rate  of  energy
transfer across the  envelope.  When convective coupling  sets in, the
envelope becomes significantly more transparent and there is initially
an excess  of thermal energy,  that the  star must radiate  away.  The
release of this excess energy delays  for a while the cooling process.
Due to  the slightly lower  adiabatic gradient in the  hydrogen layers
obtained from the \citet{Saea95} EOS (differences $\approx$10\% in the
deeper hydrogen layers), the convective envelope is more extended, and
convective coupling sets in earlier  compared to calculations with the
\citet{MM79} EOS  in the envelope.  This explains the bump  in $\Delta
t/t$ seen  around $\log(L/L_{\sun})\sim -4.2$ for  the comparison with
the standard imputs;  the bump disappears in the comparison  at step 1
(compare    the     dash-dotted    with    the    solid     line    in
Fig.~\ref{FigpureCstd})  when the  same  EOS  is employed  everywhere.
Overall, when the two sets of calculations employ the same EOS for the
whole  structures, there  are still  differences of  $\pm$10\% in  the
cooling times. Now, {\tt BaSTI} models cool systematically faster than
{\tt  LPCODE} ones  below $\log(L/L_{\sun})\sim  -2.0$, and  slower at
higher luminosities.

Employing also the same electron  conduction opacities (dashed line in
Fig.~\ref{FigpureC}, that overlaps with the solid line at luminosities
above  $\log(L/L_{\sun})\sim  -3.5$)  makes  the  cooling  times  much
closer, highlighting the  major role played by  the different electron
conduction tables  employed in  the two  codes.  The  relevant regions
(within the models) where the choice of the conduction opacities makes
a  difference  are  the  carbon   core  at  high  luminosities  (above
$\log(L/L_{\sun})  \sim  -2$),  the helium  envelope  at  intermediate
luminosites (between $\log(L/L_{\sun})  \sim -2$ and $\log(L/L_{\sun})
\sim  -4.0$), and  the hydrogen  envelope at  low luminosities  (below
$\log(L/L_{\sun})  \sim  -4.0$).   The details  of  the  opacity
  differences and their impact on the cooling timescales are discussed
  in the online appendix.

After  step  2  the  larger   disagreement  is  now  circumscribed  at
luminosities above  $\log(L/L_{\sun})\sim -1.5$  -- where the  sign of
${\rm \Delta  t}$ is  reversed compared  to the  previous step  -- and
below $\log(L/L_{\sun})\sim  -3.5$.  This latter  discrepancy vanishes
once boundary  conditions from the same  model atmosphere calculations
are  employed in  both {\tt  BaSTI} and  {\tt LPCODE  } models  (solid
line).  As  mentioned before,  in this case  we are  circumventing the
difference of low-temperature opacities adopted  in the two codes.  It
is also important  to notice that when boundary  conditions from model
atmospheres  are employed  and  matched at  our  chosen optical  depth
$\tau$=25,  the underlying  convective  envelope  (when convection  is
present) is always adiabatic. The superadiabatic layers are located at
lower optical depths, and the white dwarf model becomes insensitive to
the  choice  of the  mixing  length  value  in the  stellar  evolution
calculations.  In this case the  superadiabatic layers are included in
the model  atmosphere calculations, and the  superadiabatic convection
treatment will  affect the evolutionary model  indirectly, through the
surface  boundary condition.   From this  point of  view, very  recent
advances in  3D radiation  hydrodynamics white dwarf  model atmosphere
calculations \citep{tl:13}, hold the  promise to finally eliminate the
uncertainty  in  the  cooling  evolution   due  to  the  treatment  of
superadiabatic convection.

At the end of  this final step, when the input  physics is exactly the
same in both  codes, cooling times agree  within $\sim$2\% everywhere,
but in the region above  $\log(L/L_{\sun})\sim -1.5$ (absolute ages of
the  order  of  $\approx$10~Myr  or less),  where  differences  up  to
$\sim$8\% do still appear. As  we will discuss later, this discrepancy
is  due  to the  different  physical  conditions  of the  first  model
converged at the beginning of the cooling sequence, that are erased by
the end  of the early  phase (first  $\approx 10^8$~yr of  cooling) of
efficient  neutrino  energy  losses (the  so-called  neutrino  cooling
phase).   We  want  to  mention   that  these  discrepancies  at  high
luminosities still remain, even when we use exactly the same numerical
routine to compute neutrino emission rates.

\begin{figure}
\centering
\includegraphics[width=\columnwidth]{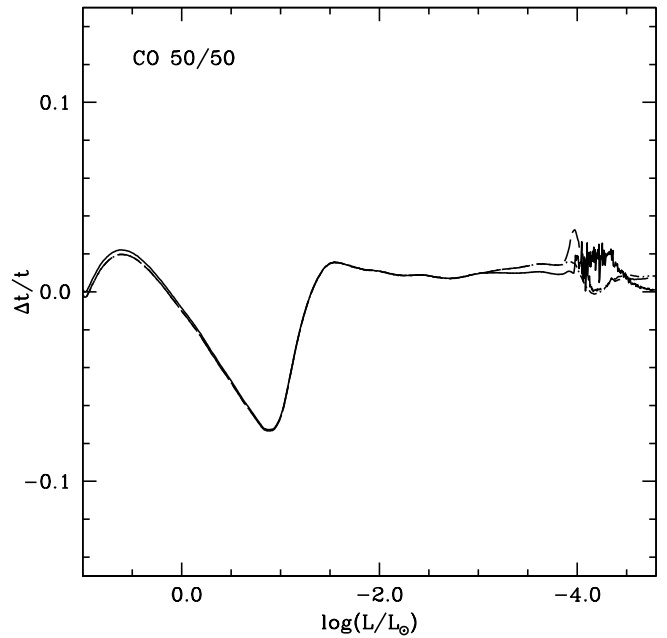}
\caption{Relative age  difference $\Delta  t/t=(t(_{\rm BaSTI})-t_{\rm
  LPCODE})/t_{\rm BaSTI}$  as a  function of luminosity  for sequences
  with cores made of a 50/50 carbon-oxygen mixture.  The input physics
  in both codes  is exactly the same,  as specified at step  3 for the
  pure  carbon  sequences.   Dash-dotted,   dashed,  and  solid  lines
  correspond to calculations without latent heat and phase separation,
  with latent heat and no phase  separation, and with both latent heat
  and phase separation, respectively.}
\label{FigCOage}
\end{figure}

A second group of cooling models, for a white dwarf with the same mass
and  envelope  stratification  but  now  a  50/50  carbon-oxygen  core
composition  (by mass),  have been  then calculated  with both  codes,
using consistent  input physics  -- as  at step 3  of the  pure carbon
models.   The purpose  of this  group  of calculations  is to  compare
results with a mixed carbon-oxygen  core composition and also with the
inclusion  of latent  heat release  -- we  adopt in  both codes  $0.77
k_{\rm  B}T$  per  crystallized  ion  --  and  phase  separation  upon
crystallization.  The test has proceeded in three steps:

\begin{enumerate}
\item  Calculations   without  release   of  latent  heat   and  phase
  separation.
\item  Calculations  with  latent   heat  release  but  without  phase
  separation.
\item Calculations with both latent heat release and phase separation.
\end{enumerate}

Both codes include  phase separation by considering in  the release of
energy per gram of crystallized  matter an extra term given by Eq.~(2)
in \citet{isern}, namely
\begin{equation}
\epsilon_{\rm g}=-\Delta X_0
\Bigg[\Big(\frac{\partial E}{\partial X_0}\Big)_{M_{\rm s}}-
\Big\langle\frac{\partial E}{\partial X_0}\Big\rangle\Bigg]
\end{equation}
where $\Delta  X_0=X_0^{\rm sol}-X_0^{\rm  liq}$ is the  difference of
oxygen  abundance   between  the  solid   and  liquid  phase   in  the
crystallizing layer  -- we  employed in  both codes  the carbon-oxygen
phase diagram of \citet{sc} -- and $E$ is the internal energy per unit
mass.  The first term represents the energy released in the layer that
is crystallizing, due  to the increase of the  oxygen abundance caused
by phase  separation, whereas the  second term represents  the average
amount of  energy absorbed in  the convective layers that  appear just
above  the  crystallization  front,  as a  consequence  of  the  local
decrease  of   the  oxygen   abundance.   The   derivative  $(\partial
E/\partial X_0)_{V,T}$ is determined layer-by-layer employing Eqs.~(6)
and (7) in \citet{isern}.

The results of  the comparisons at step 1  (displayed as a dash-dotted
line),   2  (dashed   line)  and   3   (solid  line)   are  shown   in
Fig.~\ref{FigCOage}.  The cooling  age differences between {\tt BaSTI}
and  {\tt LPCODE}  models  obtained for  these  three different  steps
overlap  at luminosities  brighter  than $\log(L/L_{\sun})\sim  -4.0$,
that marks  the onset of  crystallization.  In quantitative  terms, at
luminosities  above $\log(L/L_{\sun})\sim -1.5$,  relative differences
are slightly smaller than in the case of pure carbon models, but still
appreciable,  due again  to differences  in the  initial model  at the
start  of the cooling  sequence. At  lower luminosities,  relative age
differences are  essentially the  same as in  case of the  pure carbon
models with consistent input physics.  The inclusion of the release of
latent  heat  (dashed  line)   in  these  calculations  and  of  phase
separation (solid  line) do  not alter substantially  the quantitative
result.  There  are some  oscillations or spikes  in the  behaviour of
$\Delta  t$ with luminosity  during the  crystallization of  the core,
that can be ascribed to  the different numerical implementation of the
latent heat and  energy release due to phase  separation.  Both codes,
for  reasons of  numerical stability,  distribute this  energy release
over a narrow mass interval  around the crystallization front, and the
different implementations  of this  mechanism cause the  narrow spikes
seen in  $\Delta t$ at low  luminosities.  On the  whole, relative age
differences  are again  very small,  within $\sim$2\%  at luminosities
below $\log(L/L_{\sun})\sim -1.5$.

\begin{figure}
\centering
\includegraphics[width=\columnwidth]{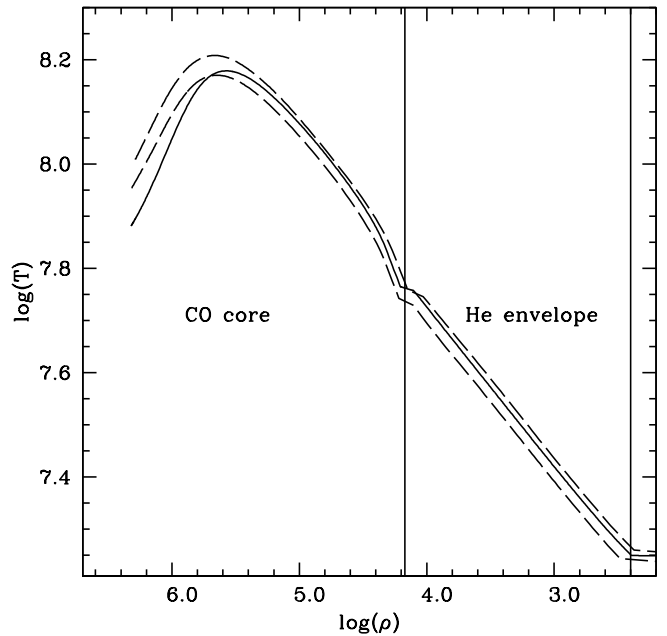}
   \caption{Temperature-density  stratification  in  two  {\tt  BaSTI}
     models (dashed lines) and one  {\tt LPCODE} model (solid line) at
     the top  of the cooling track  for the sequence in  which a 50/50
     carbon-oxygen mixture is adopted. See text for details.}
\label{Figfirstmodel}
\end{figure}

  The relative differences of the total radius $R$ are within 0.5\%
  and differences  in central temperature  $T_{\rm c}$ are  within 1\%
  for  luminosities  below  $\log(L/L_{\sun})\sim -1.5$  ({\tt  BaSTI}
  models displaying  larger $R$  and $T_{\rm  c}$), independent  of the
  inclusion of  the release  of latent  heat and/or  phase separation.
  Differences in both radius and central temperature increase steadily
  towards higher luminosities, reaching  $\sim$20\% in $T_{\rm c}$ and
  $\sim$ 3\%  in $R$ at  $\log(L/L_{\sun})\sim 1.0$.  This  is another
  consequence  of   the  different   thermal  stratification   of  the
  structures at the top of the  cooling sequence, with the {\tt BaSTI}
  models  initially   hotter  and  less  degenerate   in  the  central
  regions. Only at the end of  the neutrino cooling phase both $T_{\rm
    c}$ and $R$  converge to approximately the same values  in the two
  calculations.  Fig.~\ref{Figfirstmodel} compares the stratification
of the temperature  and density in the core of  two {\tt BaSTI} models
with luminosities equal  to, respectively, $\log(L/L_{\sun})=1.03$ and
0.82, and  a {\tt  LPCODE} model with  $\log(L/L_{\sun})=0.96$.  These
models correspond  to the $0.55\,M_{\sun}$  carbon-oxygen calculations
with the same input physics in  both codes.  In principle, if the {\tt
  BaSTI} and {\tt LPCODE} initial models were identical, the structure
of  the  {\tt LPCODE}  model  in  Fig.~\ref{Figfirstmodel} should  lie
between the  two {\tt  BaSTI} results, for  its surface  luminosity is
intermediate between the  two {\tt BaSTI} models.   Instead, the inner
part of the core (when $\log\rho > 5.6$) of both {\tt BaSTI} models is
hotter  than  the  {\tt  LPCODE}  one, a  consequence  of  the  higher
temperatures in  the core  of the  first structure at  the top  of the
cooling sequence.   This is  an important reminder  that when  sets of
white dwarf models  are calculated by {\sl  artificially} starting the
cooling sequence, with pre-determined (i.e., not derived from the full
progenitor evolution) carbon-oxygen  and envelope stratifications, the
technical details  of how  the first  white dwarf  model is  built can
induce appreciable differences in the  evolution before the end of the
neutrino cooling phase.  Our comparison provides for the first time an
order-of-magnitude estimate of these uncertainties.

\section{Conclusions}

We  have performed  the  first consistent  comparison  of white  dwarf
cooling  times determined  from two  widely used,  independent stellar
evolution codes ({\tt BaSTI} and  {\tt LPCODE}) employing {\sl exactly
the  same input  physics}.   Cooling age  differences determined  from
these  comparisons  represent the  maximum  possible  accuracy in  the
current estimates of white dwarf  cooling times, arising only from the
different implementations of the stellar evolution equations and input
physics.

We have first considered a  $0.55\, M_{\sun}$ model with a pure carbon
core, and pure hydrogen layers with mass fraction $q_{\rm H} = 10^{-4}
M_{\rm  WD}$ on top  of a  pure helium  buffer of  mass $q_{\rm  He} =
10^{-2}  M_{\rm  WD}$.   At  this  stage,  latent  heat  release  upon
crystallization  has  been neglected.   When  the  same radiative  and
electron  conduction  opacities  --  namely, OPAL  opacities  and  the
conductive   opacity  of   \citet{Caea07}  --   photospheric  boundary
conditions  --  from  the  model atmospheres  by  \citet{rohrmann}  --
neutrino  energy loss rates  \citep{Itoh2,Haft}, and  EOS \citep{MM79}
are employed, cooling  times obtained from the two  codes agree within
$\sim2\%$ at  all luminosities, with the exclusion  of models brighter
than $\log(L/L_{\sun})\sim  -1.5$, where differences up  to $\sim 8\%$
appear,  due to  the different  thermal structure  of the  first white
dwarf structure converged at the beginning of the cooling sequence.

We  have then  considered a  $0.55\,  M_{\sun}$ model  with a  uniform
carbon-oxygen stratification (50/50 by mass) in the core, and the same
hydrogen and helium envelope. We  have calculated models with the same
input physics listed above, without taking into account the release of
latent heat and  disregarding phase separation during crystallization,
including  only the release  of latent  heat, and  finally considering
both  contributions, using  the phase  diagram of  \citet{sc}  and the
treatment of \citet{isern1,isern} for  the energy release due to phase
separation.  Relative  age differences  are always very  small, within
$\sim  2\%$ at luminosities  below $\log(L/L_{\sun})\sim  -1.5$.  This
$2\%$ difference is smaller  than the current uncertainties in cooling
times due to uncertainties  in the white dwarf chemical stratification
-- see, e.g., \citet{salarisWD2}.   At higher luminosities differences
up to  $\sim 8\%$ are found  again, due to differences  in the thermal
stratification of the starting model.

As already mentioned in the introduction, to broaden the scope of
  this  study  by  facilitating   comparisons  with  additional  codes
  employed by the  white dwarf community, we provide  tables with the
{\tt BaSTI} cooling tracks for  the three carbon-oxygen sequences with
and without considering the release  of latent heat and/or considering
phase separation, and the sequence  with a carbon core without release
of latent heat, all calculated with the input physics listed above. We
also provide  the routine for  the EOS  of \citet{MM79} and  the phase
diagram of  \citet{sc} as tabulated  in our codes.  All  other physics
ingredients are  publicly available  and most of  them widely  used in
modern white dwarf calculations.

Before closing, we wish  to restate that the goal of  this study is to
compare results from independent white  dwarf cooling codes when using
the same input  physics, not to provide up-to-date  white dwarf models
to  compare   with  observations.   Specifically,  and   for  ease  of
implementation,  we  use in  our  common  physics  inputs the  EOS  of
\citet{MM79}, which is a simplification  of the EOS implemented in our
respective  codes.   As already  mentioned  in  the previous  section,
switching   from  this   EOS  to   that  of   \citet{Saea95}  in   the
hydrogen-helium  envelopes, provides  the  same  cooling times  within
$\sim  1\%$, at  luminosities brighter  than the  onset of  convective
coupling.   However,  the situation  is  different  for the  core,  as
already discussed on the basis of a comparison of the specific heat in
the  pure  carbon models.   As  an  additionaly  test using  the  {\tt
LPCODE},  we have  performed calculations  switching from  the EOS  of
\citet{MM79} to that of \citet{segretain}  in a model with pure carbon
core --  the model  at step 3  of the carbon  core comparisons  in the
previous section.  Cooling times are  reduced by up to $\sim 20-30\%$,
mainly during the crystallization phase,  when the EOS of \citet{MM79}
is    employed,    consistent    with    the    results    shown    in
Fig.~\ref{FigpureCstd}.  It is therefore  clear that this EOS provides
sizably   shorter  cooling   times   compared  to   more  modern   EOS
calculations, when employed in the degenerate carbon-oxygen cores (but
this EOS  is still  accurate for the  hydrogen and  helium envelopes).
However,  given that  what  matters  here is  the  {\sl difference  of
cooling times predicted by independent  evolutionary codes for a fixed
set  of  input  physics},  a   potential  inadequacy  of  the  EOS  of
\citet{MM79} compared to more modern EOS calculations is irrelevant in
this context, and does not affect both the outcome and the validity of
these tests.

\begin{acknowledgements}
We thank Marcelo  Miller Bertolami for helpful  discussions about some
results from LPCODE calculations.  Part  of this work was supported by
AGENCIA  through the  Programa  de  Modernizaci\'on Tecnol\'ogica  BID
1728/OC  -AR, by  PIP 112-200801-00940  grant from  CONICET, by  MCINN
grant AYA2011--23102, by the ESF EUROCORES Program EuroGENESIS (MICINN
grant EUI2009-04170),  by the European  Union FEDER funds, and  by the
AGAUR.
\end{acknowledgements}

\bibliographystyle{aa}

\Online

\begin{appendix} 

\section{Codes and input physics}

The {\tt BaSTI} code for  white dwarf evolutionary calculations solves
the four  equations describing the structure  of a star by  applying a
Raphson-Newton method,  following the  techniques described  in, e.g.,
\citet{kww:13}. The  independent variable  is the  mass $M_r$  and the
independent  variables are  radius ($r$),  pressure ($P$),  luminosity
($l$) and temperature ($T$).  The  model structure is divided into the
interior, and an external layer whose  thickness can be chosen. In our
calculations  we fixed  the mass  fraction of  this external  layer to
$q=10^{-6}$, two orders of magnitude smaller than the thickness of the
hydrogen envelope.  This extremely thin  external layer is modelled by
integrating the equations $dR$/$dP$, $dM$/$dP$ and $dT$/$dP$, with $P$
as independent variable -- considering  the luminosity $l$ constant --
by means of a fourth-order  Runge-Kutta method.  The starting point is
the bottom of the atmosphere, whose  pressure is provided either by an
Eddington  $T(\tau)$  integration,  or from  detailed  non-grey  model
atmospheres -- see \citet{salarisWD2}.   This integration provides the
outer boundary for  the Newton-Raphson integration of the  full set of
equations with  a centred scheme,  to model the  rest of the  star. In
these  white dwarf  calculations  the mass  distribution  of the  mesh
points has been set by the requirement that $R$, $l$, $P$, $T$ and $M$
do not  vary by more than  a fixed amount  from one mesh point  to the
next one, at  the beginning of the calculations.  The  total number of
mesh points in the models discussed here is $\sim$1000.  The time step
is set by the requirement that $R$,  $l$, $P$ and $T$ do not vary from
one  model to  the next,  at  each mesh  point, by  more than  $\delta
R/R=0.01$,   $\delta  l/l=0.02$,   $\delta   P/P=0.05$,  and   $\delta
T/T$=0.05.

The standard input physics includes for the carbon-oxygen core the EOS
of \citet{straniero88} in the gaseous  phase, while for the liquid and
solid phases  the detailed EOS  of \citet{segretain} is used.   As for
the  envelope H  and He  regions,  the results  of \citet{Saea95}  are
employed, supplemented at the highest densities by an EOS for H and He
using  the  prescriptions  of \citet{segretain}.   Crystallization  is
considered to occur at $\Gamma=180$,  where $\Gamma$ is the plasma ion
coupling parameter.  The associated release  of latent heat is assumed
to be equal  to $0.77 k_{\rm B}  T$ per ion ($k_{\rm  B}$ denoting the
Boltzmann constant).

The  additional  energy  release  due   to  phase  separation  of  the
carbon-oxygen  mixture  upon  crystallization  is  computed  following
\citet{isern1}  and \citet{isern}.   Neutrino energy  losses are  from
\citet{Itoh2}  and  \citet{Haft}  for plasma-neutrino  emission.   The
conductive  opacities  of  \citet{Itoh1} and  \citet{Mitake}  for  the
liquid and solid phases are adopted. For the range of temperatures and
densities not covered  by the previous results,  the conductivities by
\citet{HL69} are  instead used. OPAL radiative  opacities \citep{opal}
with $Z=0$ are used for $T>6000$~K in the He and H envelopes. In the H
envelope, and  for the temperatures  and densities not covered  by the
OPAL  tables, Rosseland  mean  opacities come  from the  monochromatic
opacities  of  \citet{sj99}.   EOS  and  opacity  tables  for  various
carbon-oxygen ratios  are employed, and the  interpolation in chemical
composition  is linear  in the  carbon abundance.   At fixed  chemical
composition the opacity  interpolation is cubic in $\log  T$ and $\log
\rho$, whilst  the EOS interpolation is  linear in $\log T$  and $\log
P$.  Convection  is treated with the  standard mixing-length formalism
by \citet{bv:58} with mixing length $\alpha$=1.5.

{\tt LPCODE} is based on the same standard method to solve the stellar
structure equations.  Three envelope integrations from the photosphere
inward to a  fitting outer mass fraction  ($q=10^{-10}$) are performed
to specify the outer boundary conditions.  The independent variable is
$\xi= \ln (1-M_r/M_*)$  and the dependent variables are  $R$, $P$, $l$
and  $T$.  The  following change  of variables  is considered  in {\tt
LPCODE}:

\begin{eqnarray}
\theta^{(n+1)} &=& \theta^{(n)} + \ln{\left(  1  +  u_\theta \right)} \nonumber \\
p^{(n+1)} &=& p^{(n)} + \ln{\left(  1  +  u_p \right)} \nonumber \\
x^{(n+1)} &=& x^{(n)} + \ln{\left(  1  +  u_x \right)} \nonumber \\
l^{(n+1)} &=& l^{(n)} + u_l
\end{eqnarray}

\noindent with $u_\theta$, $u_p$, $u_x$ and $u_l$ being the quantities
to be iterated  that are given by $u_\theta=  \Delta T/T^{(n)}$, $u_p=
\Delta  P/P^{(n)}$,  $u_x=  \Delta  R/R^{(n)}$  and  $u_l=  l$,  where
superscripts  $n$ and  $n+1$  denote  the beginning  and  end of  time
interval.  Here, $\theta=$ ln $T$, $x=$ ln $R$ and $p=$ ln $P$.  Thus,
the Newton-Raphson iteration  scheme is applied to  the differences in
the luminosity, pressure, temperature  and radius between the previous
and the computed  model. For the white dwarf regime, it is used a centred
scheme for the  equations of stellar structure  and evolution.  Models
are divided into approximately 1000-1500 mesh points.

The {\tt  LPCODE} employs the  equation of state  of \citet{segretain}
for   the   high-density  regime   (above   a   density  of   $8\times
10^2$~g/cm$^3$) complemented  with an updated version  of the equation
of  state  of  \citet{MM79}  for the  low-density  regime.   Radiative
opacities above  11\,000~K and  neutrino energy losses  are as  in the
{\tt  BaSTI}  code,  while  for temperatures  below  8000~K  radiative
opacities from the AESOPUS database  \citep{MA08} are employed, and in
the  intermediate regime  an  interpolation between  OPAL and  AESOPUS
opacities  is  performed.   OPAL radiative  opacities  are  calculated
directly  from the  interpolation routine  provided by  OPAL (version:
Arnold Boothroyd,  April 27, 2001). Arbitrary  hydrogen abundances and
arbitrary amounts of  excess carbon and oxygen are  always allowed. In
this paper, $Z=0$ is considered. The routine performs up to 6-variable
interpolation to get the opacity at the given composition, temperature
and density.

Electron conduction opacities are taken from \citet{Caea07}.  Neutrino
energy  losses, the  release of  latent heat  and carbon-oxygen  phase
separation upon crystallization are treated in  the same way as in the
{\tt  BaSTI}  code.   Latent  heat is  included  self-consistenly  and
locally coupled to the full set of equations of stellar evolution, and
is  calculated  at  each  iteration  during  the  convergence  of  the
model. The contribution is distributed  over a small mass range around
the crystallization front.  Outer boundary conditions are derived from
non-grey model atmospheres \citep{rohrmann},  or alternatively, from a
standard Eddington $T(\tau)$ relation.  Convection is treated with the
standard  \citet{bv:58}  mixing  length formalism  and  mixing  length
$\alpha=1.61$.

\end{appendix}

\begin{appendix} 

\section{The role of electron conduction opacities}

\begin{figure}
\centering
\includegraphics[width=\columnwidth]{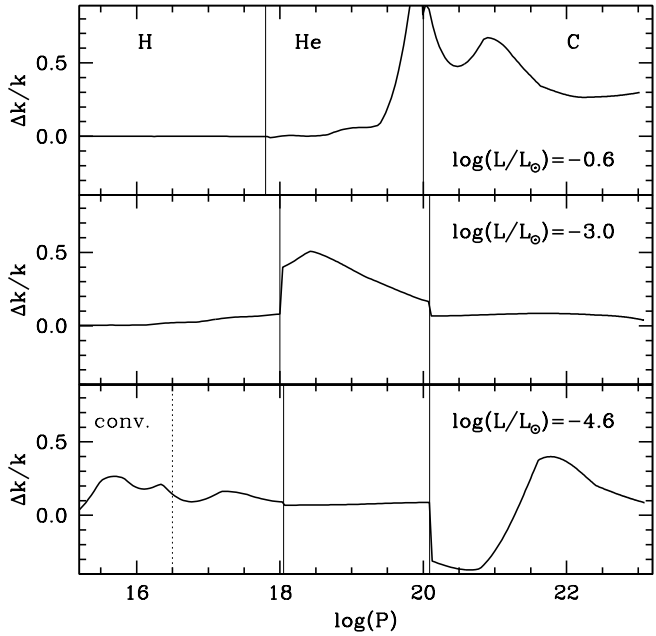}
\caption{Relative difference of the stellar  opacity $\Delta k/ k $ as
  a  function of  gas  pressure,  at fixed  $T$,  $\rho$ and  chemical
  stratifications  taken from  three models  calculated with  the {\tt
  BaSTI} code at step 2, with  the labelled luminosities (see text for
  details).  The H, He and C regions, as well as the lower boundary of
  surface convection, are marked.}
\label{Figcond}
\end{figure}

As mentioned  in our  comparison of carbon  core models,  the relevant
regions  where  the  choice  of   the  conduction  opacities  makes  a
difference   are  the   carbon  core   at  high   luminosities  (above
$\log(L/L_{\sun})  \sim  -2$),  the helium  envelope  at  intermediate
luminosites (between $\log(L/L_{\sun})  \sim -2$ and $\log(L/L_{\sun})
\sim  -4.0$), and  the hydrogen  envelope at  low luminosities  (below
$\log(L/L_{\sun})  \sim  -4.0$).   Figure~\ref{Figcond}  displays  the
difference  of the  total opacity  at fixed  $T$, $\rho$  and chemical
stratifications  (from the  {\tt BaSTI}  calculations at  step 2)  for
three different  luminosities.  The  difference is between  the values
obtained employing the \citet{Caea07}  conduction opacities, minus the
results obtained from the combination of \citet{Itoh1}, \citet{Mitake}
and \citet{HL69} conduction  opacities used as standard  input in {\tt
BaSTI}  calculations.  As  shown  by Fig.~\ref{Figcond},  in the  high
luminosity regime electron conduction  opacities by \citet{Caea07} are
higher mainly in  the carbon core, and this turns  out to increase the
neutrino  energy losses,  causing  a faster  cooling. This  connection
between increased  opacities in  the core of  bright, hot  models, and
increased  neutrino   emission  has  been  verified   by  a  numerical
experiment, where we have calculated a cooling sequence by artifically
enhancing the opacity of the carbon core at high luminosities.

At intermediate  luminosities it is  the opacity of the  helium layers
that increases,  when employing  \citet{Caea07} results. In  this case
the effect is a slowing down of the cooling, due to the higher opacity
of the helium  envelope. At low luminosities it is  instead the higher
opacity of the  degenerate hydrogen layers that has a  major effect on
the cooling speed.   We also find large variations of  the carbon core
opacity (see Fig.~\ref{Figcond}), that  are however irrelevant in this
regime, because the opacity of the core is anyway extremely small, due
to  the high  degree of  degeneracy  of the  core.  The  effect of  an
increased opacity  of the hydrogen  layers at low  luminosity actually
speeds-up  the  cooling,  as  we   have  confirmed  with  a  numerical
experiment where  we have computed  a model by  enhancing artificially
the opacity  of the  degenerate layers of  the hydrogen  envelope. The
reason  for  this  counterintuitive  result  is  the  slightly  deeper
extension  of  the hydrogen  convective  envelope  in this  luminosity
regime.  In fact, after convective coupling has been established -- at
$\log(L/L_{\sun}) \sim  -4.0$ in these  models -- the  cooling becomes
faster  compared  to the  case  of  no  coupling,  because of  a  more
efficient  energy  transport.  The   energy  transport  efficiency  is
increased when convection extends  deeper into the degenerate hydrogen
layers, thus increasing  also the cooling speed.   These three effects
explain the difference  between the $\Delta t/t$  values determined at
step 1 and 2 of our pure carbon core model comparisons.

\end{appendix}

\end{document}